\documentclass[
reprint,
twocolumn,
aps,
prr,
citeautoscript,
showpacs,
showkeys,
groupedaddress,
superscriptaddress,
twocolumn
amsmath,
amssymb,
]{revtex4-2}

\usepackage{graphicx}
\usepackage{dcolumn}
\usepackage{bm}

\usepackage[hidelinks]{hyperref}
\hypersetup{
    colorlinks,
    linkcolor={blue!50!black},
    citecolor={blue!50!black},
    urlcolor={blue!80!black}
}

\usepackage{siunitx}
\usepackage{xcolor}
\usepackage{braket}
\usepackage{enumitem}

\usepackage{mathtools}
\newcommand{\disorderstrength}[0]{\ensuremath{\langle V \rangle}}
\newcommand{\boltzmann}{k_\mathrm{B}}
\newcommand{\avg}[1]{\langle #1 \rangle}

\begin{document}

\title{What can we learn from diffusion about Anderson localization\\ of a degenerate Fermi gas?}

\author{Sian Barbosa}
\affiliation{Department of Physics and Research Center OPTIMAS, RPTU Kaiserslautern-Landau, 67663 Kaiserslautern, Germany}

\author{Maximilian Kiefer-Emmanouilidis}
\affiliation{Department of Physics and Research Center OPTIMAS, RPTU Kaiserslautern-Landau, 67663 Kaiserslautern, Germany}
\affiliation{Department of Computer Science, RPTU Kaiserslautern-Landau, 67663 Kaiserslautern, Germany}
\affiliation{Embedded Intelligence, German Research Centre for Artificial Intelligence, 67663 Kaiserslautern, Germany}

\author{Felix Lang}
\affiliation{Department of Physics and Research Center OPTIMAS, RPTU Kaiserslautern-Landau, 67663 Kaiserslautern, Germany}

\author{Jennifer Koch}
\affiliation{Department of Physics and Research Center OPTIMAS, RPTU Kaiserslautern-Landau, 67663 Kaiserslautern, Germany}

\author{Artur Widera}
\email[]{widera@rptu.de}
\affiliation{Department of Physics and Research Center OPTIMAS, RPTU Kaiserslautern-Landau, 67663 Kaiserslautern, Germany}

\date{\today}

\begin{abstract}
Disorder can fundamentally modify the transport properties of a system. 
A striking example is Anderson localization, suppressing transport due to destructive interference of propagation paths. 
In inhomogeneous many-body systems, not all particles are localized for finite-strength disorder, and the system can become partially diffusive. 
Unraveling the intricate signatures of localization from such observed diffusion is a long-standing problem.
Here, we experimentally study a degenerate, spin-polarized Fermi gas in a disorder potential formed by an optical speckle pattern. 
We record the diffusion through the disordered potential upon release from an external confining potential. 
We compare different methods to analyze the resulting density distributions, including a new approach to capture particle dynamics by evaluating absorption-image statistics. 
Using standard observables, such as diffusion exponent and coefficient, localized fraction, or localization length, we find that some show signatures for a transition to localization above a critical disorder strength, while others show a smooth crossover to a modified diffusion regime. 
In laterally displaced disorder, we spatially resolve different transport regimes simultaneously, which allows us to extract the subdiffusion exponent expected for weak localization. 
Our work emphasizes that the transition toward localization can be investigated by closely analyzing the system's diffusion, offering ways of revealing localization effects beyond the signature of exponentially decaying density distribution. 
\end{abstract}

\maketitle

\section{Introduction}
\label{sec:intro}

In 1958, P.~W.~Anderson showed that a single electron in a disordered material will show an absence of diffusion at long times, i.e., it localizes~\cite{anderson_absence_1958, AndersonLocalization}. 
The tails of the particle's wave function or probability density profile decay exponentially in space $\avg{n(r)}\sim \mathrm{exp}(-|r|/\xi)$, where $n$ is the particle density, $r$ is the position, $\xi$ is the localization length, and $\avg{\cdot}$ is a disorder and quantum average. 
In one and two dimensions, localization in random potentials always emerges from a particle interfering with its multiple scattering paths. 
This interference is constructive only at the spatial point where the particle has been initially placed. 
In $d = 3$ dimensions, the system only localizes for sufficiently strong disorder, as not all scattering paths are expected to return to the origin. 
Thus, localization is expected only if the disorder becomes sufficiently strong~\cite{AndersonLocalization}. 
In this case, a metal-insulator transition is expected with power law divergence at a critical value $\xi(E) \sim |E - E_\mathrm{c}|^{-\nu}$ where $E$ is the energy, $E_\mathrm{c}$ is a critical energy called “mobility edge", and $\nu \approx \SI{1.58}{}$ is the critical exponent~\cite{MacKinnon1983, slevin_corrections_1999, lopez_experimental_2012}. 
Within the last 65 years, such Anderson localization (AL)~\cite{anderson_absence_1958, AndersonLocalization} has been observed in non-interacting systems of light~\cite{Wiersma1997, Scheffold1999, Stoerzer2006, schwartz_transport_2007, Sperling2013,Maret2013,Mafi2021}, ultrasound~\cite{Weaver1990, Hu2008, goicoechea_suppression_2020}, and microwaves~\cite{Dalichaouch1991, Chabanov2000}. 
In the context of ultracold atoms, AL has been intensively studied in 1D, both in the quasiperiodic lattice~\cite{roati_anderson_2008} and in the continuum~\cite{billy_direct_2008}, in 2D~\cite{white_observation_2020, shamailov_comment_2021, najafabadi_effects_2021}, 
as well as in 3D~\cite{kondov_three-dimensional_2011, jendrzejewski_three-dimensional_2012, semeghini_measurement_2015}. 

During the last years, the investigation of the Anderson metal-insulator transition has increased significantly~\cite{ZhaoSirker2020, Sierant2020AL, Suntajs2021AL} due to an active debate whether an interacting system can localize at a finite disorder strength~\cite{KieferUnanyan2, KieferUnanyan3, Suntajs2019, Suntajs2020, Abanin2021, Ghosh2022, Luitz2020, Leonard2023, Sels2021, Sels2022, sierant_challenges_2022}. 
Moreover, even in the non-interacting case, the literature seems not fully conclusive as the most striking feature, the exponential tails of Anderson-localized particles, can simply be an artifact from an energy-dependent diffusion coefficient $D(E)$~\cite{shapiro_expansion_2007, shapiro_cold_2012, muller_comment_2014}. 
In particular, this is a problem for fermionic systems, where a broad distribution of $D(E)$ due to Pauli exclusion is present. 
Since the observation of an exponential decay of the wavefunction as a smoking gun for AL is not sufficient, other signatures for localization from experimentally accessible observables are required. 

A common technique to study transport phenomena is via a quantum quench~\cite{billy_direct_2008, roati_anderson_2008, robert-de-saint-vincent_anisotropic_2010, jendrzejewski_three-dimensional_2012, kondov_three-dimensional_2011}. 
Here, an initial equilibrium state is prepared in a harmonic trap, usually at a low temperature, and then quenched into a disorder potential while simultaneously extinguishing the trap. 
Classically, one expects the particles above the percolation threshold to undergo standard Brownian diffusion when expanding through such a disordered potential landscape. 
Due to quantum interference effects leading to localization, however, anomalous subdiffusion is expected for particles close to the mobility edge~\cite{scholak_spectral_2014, shapiro_cold_2012} 
while the diffusion coefficient approaches the Heisenberg-limited ratio of the reduced Planck constant and the particle mass, the “quantum of diffusion" $\hbar / m$~\cite{kuhn_coherent_2007, jendrzejewski_three-dimensional_2012, semeghini_measurement_2015, muller_comment_2014, shapiro_cold_2012, patel_universal_2020, sommer_universal_2011, enss_quantum_2012}. 
Hence, the evolution of a system toward localization can be investigated by closely observing its diffusion. 

Generally, different regimes of anomalous diffusion are distinguished by the exponent $\alpha$ of the growing mean squared displacement (MSD) of a particle from its initial position $r(t=0)$,
\begin{equation}
\label{eq:msd_anomalous_diffusion}
    \mathrm{MSD}(r(t))\equiv \langle |r(t) - r(0)|^2 \rangle = 2d D_\alpha t^\alpha,
\end{equation}
which evolves over time $t$ and anomalous diffusion coefficient $D_\alpha$~\cite{munoz-gil_objective_2021, metzler_modelling_2022, metzler_random_2000, mangalam_ergodic_2023}. 
For $\alpha = 1$, the system undergoes normal diffusion, subdiffusion for $\alpha < 1$, and superdiffusion for $\alpha > 1$. 
Another special case, $\alpha = 2$, is called ballistic and describes the free expansion without any medium while also occurring in non-trivial cases~\cite{metzler_random_2000, levi_hyper-transport_2012, munoz-gil_objective_2021}. 

Here, we study the expansion of a non-interacting Fermi gas into a disordered optical waveguide. 
Recording the dynamics of the density distribution upon releasing the system from the trap allows us to observe the diffusive expansion and deviations from normal diffusion, which we attribute to signatures of Anderson localization. 
We compare different methods to obtain anomalous-diffusion observables, i.e., exponent and coefficient, from absorption images taken, and we analyze them for signatures for localization.

\section{Experimental methods}
\label{sec:experimental_methods}

\begin{figure*}
    \includegraphics[width=178mm]{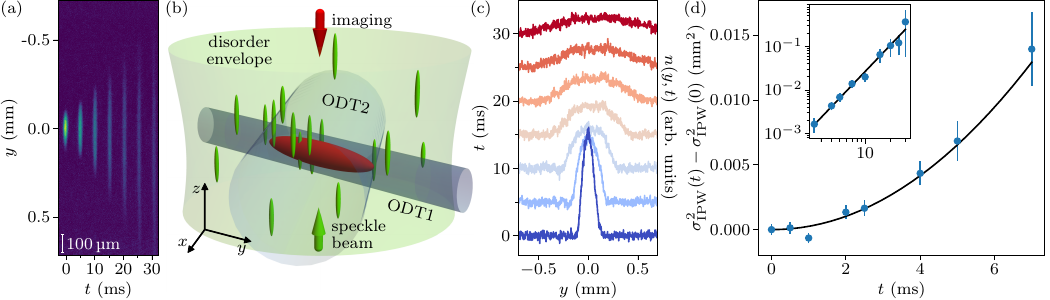}
    \caption{Schematic illustration of the experimental setup, absorption images, and evaluation of disorder-free ballistic expansion. 
    (a)~Absorption images taken at various expansion durations $t$. Fifty repetitions are averaged per image.
    (b)~Experimental setup. The atomic cloud (red ellipsoid) is initially trapped in a superposition of two optical dipole traps (blue tubes). The speckle beam (green volume and arrow), propagating along the $+z$ axis, produces randomly distributed anisotropic grains (green ellipsoids). At $t=0$, the crossed dipole trap is extinguished, and the atoms begin to expand along the $y$ axis. The cloud is imaged after a variable expansion duration $t$ with a resonant beam (red arrow) propagating along the $-z$ direction. 
    (c)~Density profiles $n(y, t)$ for equidistant time steps (bottom to top, left axis). The profiles are calculated from the absorption images shown in (a) by integrating along the $x$ axis. 
    (d)~Variance $\sigma^2_\mathrm{IPW}(t) - \sigma^2_\mathrm{IPW}(0)$ over time for short timescales, the solid line is a power-law fit with exponent $\alpha = 2.02 \, \pm \, 0.09$, see Eq.~\eqref{eq:anomalous_diffusion}. Error bars are calculated from error propagation. In the inset, the same fit with more data for longer timescales is shown in a double-logarithmic plot. 
    }
    \label{fig1}
\end{figure*}

Experimentally, we prepare a degenerate Fermi gas of typically $N \approx 8 \times 10^5$ $^6$Li atoms, polarized in the lowest-lying Zeeman substate and inside a crossed optical dipole trap (see Fig.~\ref{fig1}(b)). 
Due to the fermionic nature of the atoms, for the low temperature $T \approx \SI{100}{\nano K} \approx \SI{0.15}{} \, T_\mathrm{F}$ (with Fermi temperature $T_\mathrm{F}$) of the atomic samples, $s$-wave interactions are prohibited, while $p$-wave and higher order interactions are strongly suppressed. 
Hence, in good approximation, the atomic cloud behaves as an ideal Fermi gas~\cite{beilin_diffusion_2010, shapiro_cold_2012, top_spin-polarized_2021}. 
Initially, the atoms are trapped in an optical dipole trap (ODT), created by superposing a focused laser beam ODT1, propagating along the $y$ axis, with a secondary beam ODT2, crossing the first laser beam at an angle of 53° in the $x$-$y$ plane, see Fig.~\ref{fig1}(b). 
The resulting crossed trap has the trap frequencies $(\omega_x^\times, \omega_y^\times, \omega_z^\times) = (386, 37.8, 257) \times 2 \pi \, \SI{}{Hz}$, with the superscript~$\times$ denoting the crossed trap. 
By extinguishing ODT2 at $t=0$, the trap frequencies become $(\omega_x, \omega_y, \omega_z) = (365, 1.9, 248) \times 2\pi \, \SI{}{Hz}$. 
This happens instantaneously, i.e., at a duration smaller than $\SI{1}{\micro\second}$, much shorter than the inverse trap frequency. 
The trap geometry along the $x$ and $z$ axis does not change significantly, $\omega_{x,z} \approx \omega_{x, z}^\times$, while the potential along the $y$ axis effectively becomes flat, $\omega_y \ll \omega_y^\times$, allowing the atoms to expand along the $y$ direction. 
After a variable expansion duration $t$, we perform resonant high-intensity absorption imaging~\cite{reinaudi_strong_2007, ries_observation_2015, nagler_cloud_2020}
along the $z$ axis. 

To probe diffusion in disorder, a repulsive optical speckle disorder potential $V(x, y, z)$ composed of \SI{532}{\nano\meter} laser light and with a typical grain size $\eta_{x,y}^2 \times \eta_z = (\SI{750}{\nano\meter})^2 \times \SI{10.2}{\micro\meter}$ is quenched on at $t=0$, where $\eta_{x,y}$ and $\eta_z$ are the correlation lengths along the respective directions~\cite{nagler_cloud_2020, nagler_observing_2022, kuhn_coherent_2007}. 
We note that this three-dimensional speckle potential does not allow for classically bound states~\cite{pilati_dilute_2010, sanchez-palencia_disorder-induced_2008, delande_mobility_2014}. 
We characterize the disorder strength by its spatial average $\disorderstrength$. 
In this work, we investigate the effect of disorder strengths of up to $\disorderstrength \approx \SI{700}{\nano K} \times \boltzmann$, which can be somewhat larger than the typical Fermi energies $E_\mathrm{F} \approx \SI{550}{\nano K} \times \boltzmann$. 

For all settings, i.e., expansion duration and disorder strength, we create and image 50 realizations of the atom sample and turn the speckle diffusor plate by a small angle in between to ensure a unique disorder realization for each shot. 
These images are averaged to ensure a good disorder average~\cite{shapiro_cold_2012} and reduce the impact of noise in the evaluation. 
For this setup in particular, image noise is enhanced due to technical details (see Appendix~\ref{app:methods}) compared to our previous works, requiring more averaging for the same signal-to-noise ratio.
To analyze the expansion along the $y$ direction, we integrate the averaged images taken at time $t$ over the $x$ axis, see Fig.~\ref{fig1}(a) and (c). 
This yields the column densities $n(y, t)$ for the one-dimensional density distributions we analyze as described in Sec.~\ref{sec:diffusion_observables}. 

Finally, we use $d=1$ in the diffusion analysis as the dimensions $x, z$ are blocked from undergoing diffusion due to the harmonic confinement or, in other words, $D_{x, z} = 0$. 
However, we emphasize that the gas as well as the speckle disorder never cross into the regimes of dimensionality lower than $d=3$.

\section{Diffusion observables}
\label{sec:diffusion_observables}

In non-interacting systems, we expect particles to diffuse ballistically, normally or anomalously (subdiffusively~\cite{shapiro_cold_2012}) when the system is subjected to no disorder, weak disorder, or strong disorder, respectively. 
Ideally, diffusion is investigated by evaluating the mean squared displacement computed from trajectories of single particles traced over time~\cite{kindermann_nonergodic_2017, munoz-gil_objective_2021}. 
Since we, like the majority of cold-gas experiments, cannot physically access the individual-trajectory MSD as defined in Eq.~\eqref{eq:msd_anomalous_diffusion}, we need to restrict ourselves to observables characterizing the ensemble of many particles. 
In this section, we look at four different observables of the spatial variance $\sigma^2_i$ of the cloud that we briefly introduce and then compare each to their ability to extract the diffusion exponent $\alpha$ as well as coefficient $D_\alpha$ from
\begin{equation}
\label{eq:anomalous_diffusion}
    \sigma_i^2(t) - \sigma_i^2(0) = 2 D_\alpha t^\alpha. 
\end{equation}
Here $i$ stands for the different methods to extract the variance,
\begin{enumerate}[label=(\Alph*)]
    \item $\sigma_\mathrm{fit}^2$, the variance of a Gaussian fitted to the density profile, see Sec.~\ref{subsec:gauss_fit_variance},
    \item $\sigma_\mathrm{Var}^2$, the spatial variance of the density profile, see Sec.~\ref{subsec:distribution_variance},
    \item $\sigma_\mathrm{PR}^2$, the variance extracted from the participation ratio (PR),  see Sec.~\ref{subsec:participation_ratio}, and
    \item $\sigma_\mathrm{IPW}^2$, the variance extracted from the inverse participation width (IPW), see Sec.~\ref{subsec:ipw}.
\end{enumerate}
To find an appropriate estimate for the spatial variances of the cloud, we use a Gaussian function
\begin{equation}
\label{eq:gauss_wavefunction}
    n_\mathrm{Gauss}(y,y_0,t) = \frac{N(t)}{\sqrt{2\pi \sigma^2(t)}} \mathrm{exp}\left( - \frac{(y-y_0)^2}{2\sigma^2(t)}\right)
\end{equation}
as a control distribution representing the density profile of the experimental data (see Fig.~\ref{fig1}(c)), with atom number $N(t) = \sum_y n(y, t) \Delta y$, density $n(y, t)$ on position $y$ and $\Delta y$ the size of a pixel. 
Note that we fix $y_0 = 0$ in the following. 
Here, the variance is time-dependent, following Eq.~\eqref{eq:anomalous_diffusion} with $\alpha = 2$ for ballistic, $\alpha = 1$ for normal, and $\alpha < 1$ for anomalous subdiffusion, respectively. 

The accurate determination of the diffusion exponent remains an actively investigated topic~\cite{munoz-gil_objective_2021}. 
For the present work, we chose the following procedure. By taking the logarithm of Eq.~\eqref{eq:anomalous_diffusion}, 
\begin{equation}
\label{eq:log_anomalous_diffusion}
    \ln \left(\sigma_i^2(t) - \sigma_i^2(0) \right) = \ln \left( 2 D_\alpha t^\alpha \right) = \ln{\left( 2 D_\alpha \right)} + \alpha \ln t,
\end{equation}
the diffusion exponent $\alpha$ can be directly inferred via linear regression. 
We use the standard fit error as the uncertainty of the exponent's error. 
An example of this fit to $\sigma_\mathrm{IPW}^2(t) - \sigma_\mathrm{IPW}^2(0)$ as extracted from data from a ballistic expansion without disorder, which is discussed in more detail in Sec.~\ref{subsec:observable_performance}, is shown in Fig.~\ref{fig1}(d). 
For the anomalous diffusion coefficient, we calculate 
\begin{equation}
\label{eq:diffusion_coefficient}
    D_\alpha = \left\langle\frac{\sigma_i^2(t) - \sigma_i^2(0)}{2 t^\alpha}\right\rangle.
\end{equation}
Note that the set of values for different $t$ that are averaged is constant in time. 
For the error of $D_\alpha$, we use the standard deviation of that set. 
Even though $D_\alpha$ is already contained in the fit result as the variance-axis intercept, we decide for this option to be less dependent on fit results. 
Note that both options yield values that are equal within their ranges of uncertainties. 

Comparing the anomalous diffusion coefficient $D_\alpha$ for different $\alpha$ is not straightforward. 
Technically, it has the unit of $\mathrm{m}^2 \, \mathrm{s}^{-\alpha}$, or normalized to the diffusion quantum $\hbar/m$ as $m D_\alpha / \hbar$, the unit of $\mathrm{s}^{1-\alpha}$~\cite{metzler_random_2000, metzler_modelling_2022}. 
The alternative method is to remove the dependence on $\alpha$ by focusing on $D_1(t) = D_\alpha t^{\alpha - 1}$ which has the unit of \SI{}{\square\meter\per\second}. 
However, $D_1(t)$ will then not be a constant in time for $\alpha \neq 1$, making a direct comparison of single representative values of the diffusion coefficient for different settings with different $\alpha$ not possible. 
Hence, we decide to evaluate $D_\alpha$ as described.

\subsection{Gauss-fit variance}
\label{subsec:gauss_fit_variance}

One of the most simple and also most common methods is to fit a distribution function to the recorded density profiles~\cite{robert-de-saint-vincent_anisotropic_2010, roati_anderson_2008, derrico_quantum_2013, vilk_unravelling_2022}, see Fig.~\ref{fig2} (a), and extract values such as the variance or peak height. 
On the one hand, fitting has the advantage of approximating the raw data without other forms of data filtering or additional averaging. 
On the other hand, the results strongly depend on the fitting function and number of fit parameters. 
In particular, the number of fit parameters should be restricted to as few parameters as possible to avoid overfitting and, therefore, omitting the general properties of the dynamics~\cite{dyson_meeting_2004, mayer_drawing_2010}. 
In the case of normal diffusion, fitting Eq.~\eqref{eq:gauss_wavefunction} becomes straightforward. 
The variance $\sigma_\mathrm{fit}^2$ is then directly extracted from the fitted parameters. 
As the density profiles consist of about $10^5$ fermions, one would naively expect the central-limit theorem to be valid. 
This is, however, only the case when the diffusion coefficient is not energy-dependent. 
Since this is generally not the case in our experiment, deviations from the Gaussian density distribution will be present in the profiles, which we discuss in more detail in Appendix~\ref{app:density_profiles}. 
Still, a Gaussian fit is a good approximation for the first and second moment $\sigma^2(t)$ of the true density profile and, thus, sufficient to extract the diffusion exponent and coefficient.

\subsection{Variance of the density profile}
\label{subsec:distribution_variance}

Another common and more direct method is to compute the mean squared displacement of the distribution~\cite{jendrzejewski_three-dimensional_2012}
\begin{equation}
    \sigma_\mathrm{Var}^2(t) \coloneqq \frac{1}{N(t)} \sum_{y} n(y, t) \, (y - y_\mathrm{com})^2 \Delta y,
\end{equation}
with the center-of-mass position $y_\mathrm{com}$. 
As $y_\mathrm{com}$, we choose the cloud-peak position taken from fitting the Gaussian since it is significantly more stable against noise when compared to calculating the center-of-mass position directly as the mean of the distribution. 
Similarly, we need to omit negative $n(y, t)$ arising from our imaging system, which is calibrated to $\avg{n} = 0$ if no atoms are present.

\subsection{Participation ratio}
\label{subsec:participation_ratio}

As a complementary approach, we present the participation ratio (PR)~\cite{bell_atomic_1970, edwards_numerical_1972, schwartz_transport_2007, dikopoltsev_observation_2022, laflorencie_entanglement_2022}
\begin{equation}
    \mathrm{PR}(t) = \frac{\left(\sum_{y} n(y, t) \Delta y\right)^2}{\sum_{y} n^2(y, t) \Delta y}.
\end{equation}
Note that PR$(t)$ has the physical dimension of length here, describing the cloud width here. 
Here, the inverse PR is proportional to the probability of a particle returning to the same position after infinite time~\cite{kramer_localization_1993}.
Similarly, the cloud's displacement becomes $\mathrm{PR}(t)^2\sim D_\alpha t^\alpha$ and is expected to follow $\mathrm{PR} = 4 \xi$ in the perfectly localized case in the continuum~\cite{laflorencie_entanglement_2022}. 
Although not commonly used to investigate diffusion, we expect PR to be constrained by upper bounds of particle transport~\cite{Hartman2017}. 
To estimate the variance, we evaluate PR for the Gaussian control distribution, Eq.~\eqref{eq:gauss_wavefunction}, 
\begin{equation}
    \sigma_\mathrm{PR}^2(t) \coloneqq \frac{\mathrm{PR}^2(t)}{4 \pi}. 
\end{equation}
This quantity tends to underestimate the diffusion exponent and coefficient.

\subsection{Inverse participation width}
\label{subsec:ipw}

\begin{figure*}
    \centering
    \includegraphics[width=178mm]{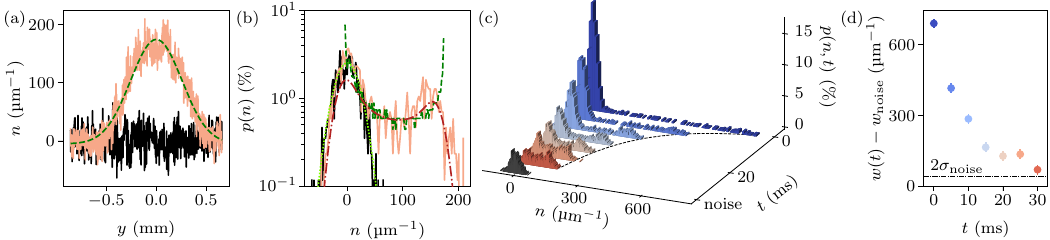}
    \caption{Density-histogram analysis.
    (a)~Noisy line-density plot (orange line) as calculated from absorption image and its Gauss fit (dark-green dashed line) as well as line density from noise measurement (black line). Here, data from the disorder-free expansion at $t=\SI{20}{\milli\second}$ are shown as a representative example. 
    (b)~Histograms of the data shown in panel (a), with the same line colors and types. Additionally, the noise histogram is fitted with a normal distribution (light-green dash-dotted line). Further, the convolution of the fit-function histogram with the noise Gauss-fit is plotted (red dash-dotted line) to visualize the convolution character of histograms. 
    (c)~Evolution of histograms over time, shown as bar plots for the same data set as shown in Fig.~\ref{fig1}. The noise histogram is additionally plotted at the front in black (labeled “noise"). The black dashed line below the decreasing histogram width is a guide to the eye. 
    (d)~Widhts $w(t)$ of all histograms shown in (c) over time, offset by noise-histogram width $w_\mathrm{noise}$. The dash-dotted line shows $2\sigma_\mathrm{noise}$ as the lower validity bound of IPW. Error bars are the standard deviation of the noise histogram. 
    }
    \label{fig2}
\end{figure*}

Access to local counting statistics~\cite{kindermann_nonergodic_2017, lukin_probing_2019, Rispoli2019, Contessi2023, Barghathi2018,Calabrese2020, kiefer-emmanouilidis_bounds_2020, KieferUnanyan4,Parez_2021,zhao_entanglement_2020} allows obtaining fundamental indicators for general particle diffusion and localization. 
Extracting these statistics is, however, not feasible in experimental setups with large amounts of particles and high densities. 
Similarly, when considering the Gaussian camera noise, we would need thousands of realizations to build a significant statistic. 
Instead of focusing on particle statistics, complementary approaches have evaluated the intensity contrast of the measured image~\cite{park_indicators_2021} as an indicator of (de)localization. 
Similarly, we decided to investigate the image histograms, which correspond to the distribution of particle densities recorded in an absorption image. 
By inspecting the normalized ordinary histogram $p(n, t)$ and the underlying distribution of densities, we gain additional knowledge to extract the diffusion details. 

We calculate $p(n, t)$ from the line density $n(y, t)$ by counting the number of densities that fall into 100 bins between the smallest and largest density recorded at time~$t$. 
We normalize $\sum_n p(n, t)=1$, such that $p(n, t)$ corresponds to a discrete probability density function (PDF) of the occurrences of densities $n$. 
Examples of density distributions and their histograms are shown in Fig.~\ref{fig2}(a,~b), where we discuss the different cases in more detail below. 

Here, we introduce a noise-robust method to extract the peak density, which we call the “inverse participation width" (IPW), a quantity based on the width of these histograms. 
From IPW, we extract $\sigma^2_\mathrm{IPW}$ and, thus, the diffusion exponent and coefficient. 
The histograms provide criteria for when IPW is a good quantity to extract the diffusion details and give an insight into how noise and IPW are related, which we discuss later in this paragraph. 
To get a more profound view of the image-histogram dynamics, we first discuss the histogram shapes and IPW itself for noise-free densities, then for the histogram contribution of camera noise alone, and finally for the combined histograms as recorded in the experiment. 

We begin with the noise-free case and a unimodal density distribution in space such as a Gaussian (see Eq.~\eqref{eq:gauss_wavefunction} and the green dashed line in Fig.~\ref{fig2}(a)). 
Then, the image histogram $p(n,t)$ shows two effectively diverging flanks at $n = 0$, no particles, and $n = n(y=0, t) = N(t) / \sqrt{2 \pi \sigma^2(t)}$ for the Gaussian function, corresponding to the peak density at time $t$ (see green dashed line in Fig.~\ref{fig2}(b)). 
Further, the kernel of the histogram, see appendix~\ref{app:histograms}, follows
\begin{align}
    p_\mathrm{Gauss}(n,t)\sim\frac{1}{n\sqrt{-\mathrm{ln}\left(\frac{n\sqrt{2\pi\sigma^2(t)} }{N(t)}\right)}}
\end{align}
for  $0 < n < n(0, t)$, the support of the kernel, which is given by the peak amplitude $n(0, t)$ of the distribution. 
This means that the full width $w(t) \coloneqq n_\mathrm{max}(t) - n_\mathrm{min}(t)$ of the histogram is equal to the peak density $n(0, t)$, which holds for any unimodal distribution in space. 
From the peak density, we can extract the Gauss variance directly as $\sigma^2(t) = N^2(t) / (2 \pi w^2(t))$. 
Note that this relation for $\sigma$ only holds in the noise-free case. 
The variance we extract from the experimental data, in which noise plays a significant part, is discussed further below. 
We consider IPW a good observable to extract $\sigma(t)$ as long as the noise-free profile of $p(n,t)$ is effectively diverging at the flanks, meaning that the decreasing width $w(t)$ captures the most relevant part of the distribution change. 
Like the other diffusion observables, IPW is not sensitive to the tails of spatial distributions. 

In the second case, we focus on the camera noise by itself. 
Experimentally, we expect noise in absorption images to result from spatial and temporal atom-number or imaging-beam-intensity fluctuations, random influences by stray light or similar factors, as well as camera noise, including shot noise, and thermal or electronic fluctuations.
To investigate the noise in the absorption images, we conduct the experimental sequence without the sample and take the average from 50 images with only noise. 
We calculate the histogram $p_\mathrm{noise}$ of the averaged image and fit it with a Gaussian function $p^\mathrm{fit}_\mathrm{noise}$, see light-green dotted line in Fig.~\ref{fig2}(b), which we find to agree very well with the noise histogram. 
Finally, we compute the width $w_\mathrm{noise}$ of the histogram and use the standard deviation $\sigma_\mathrm{noise}$ from the fit as the statistical uncertainty for $w$ and $n$. 

In the third case, we consider the histogram $p_\mathrm{data}$ of the noisy density distribution and, hence, the experimentally relevant case. 
With noise, the density distribution fluctuates, yielding a wider histogram as $n < 0$ and $n > n(0, t)$ will be observed as well. 
In fact, as is expected for the probability density function of the sum of two random variables, the combined histogram $p_\mathrm{data}$ of signal $p$ and noise $p_\mathrm{noise}$ is given by the convolution 
\begin{equation}
\label{eq:histogram_convolution}
    p_\mathrm{data}(n,t) = (p_\mathrm{noise} * p)(n) = \sum_{n^{\prime}=-\infty}^\infty p_\mathrm{noise}(n^{\prime}) p(n-n^{\prime}, t),
\end{equation}
see the orange solid line for the data histogram $p_\mathrm{data}$ as well as the red dash-dotted line for $p^\mathrm{fit}_\mathrm{noise} * p_\mathrm{Gauss}$ in Fig.~\ref{fig2}(a). 
The resulting distribution is still bimodal, where the peaks are approximately located at the effectively diverging flanks of the noiseless distribution, e.g., $p(n,t)=p_\mathrm{Gauss}(n,t)$. 
Thus, for $w(t) - w_\mathrm{noise} > 2 \sigma_\mathrm{noise}$, the resulting histogram will be bimodal~\cite{Schilling2002}, see Fig.~\ref{fig2}(c,~d). 

As the convolution near the diverging flanks resembles the convolution kernel, and hence the noise histogram $p_\mathrm{noise}$, we can approximate the noise-free peak density simply by $n(0, t) \approx w(t) - w_\mathrm{noise}$. 
With that, we define the noise-corrected inverse participation width as
\begin{equation}
\label{eq:ipw_def}
    \mathrm{IPW}(t) \coloneqq \frac{N(t)}{w(t) - w_\mathrm{noise}}
\end{equation}
and the resulting observable for the variance is then 
\begin{equation}
\label{eq:ipw}
    \sigma^2_\mathrm{IPW}(t) \coloneqq \frac{1}{2\pi} \mathrm{IPW}^2(t). 
\end{equation}
Note that the diffusion exponent $\alpha$ can be directly inferred from IPW$^2(t)$ without any specific assumptions about the density profile except for its spatial unimodality. 

We note that IPW in itself does not require any calculation of the histogram $p(n, t)$, as the width $w(t)$ can be extracted directly from the image. 
Since the validity and properties of IPW are derived directly from the image histograms and the underlying statistics, we chose to motivate the observable with the histogram analysis. 
We further state that the noise correction is intrinsic in the sense that the statistical properties of both the data and the system are exploited to reduce the effects of noise rather than trying to remove the noise from the data itself. 
For all other observables, noise reduction methods such as smoothing or wavelet decomposition are applied directly to the absorption images, significantly affecting them and potentially the extracted result. 

Finally, we note that our efforts to solve the deconvolution problem of extracting $p$ from Eq.~\eqref{eq:histogram_convolution} have not produced satisfactory results for the experimental data. 
However, the application of machine learning to this problem could be a good prospect that we will explore in a future work.

\subsection{Comparison of observable performance}
\label{subsec:observable_performance}

\begin{figure*}
    \centering
    \includegraphics[width=178mm]{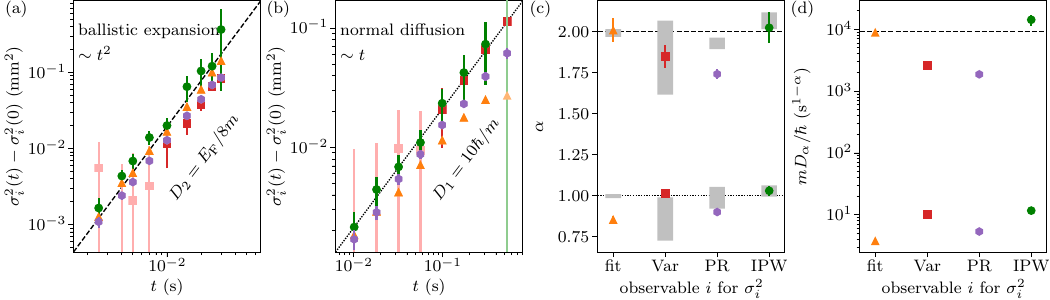}
    \caption{Comparison of observable performance. 
    Shown are the values for $\sigma_i^2(t) - \sigma_i^2(0)$ in a double-logarithmic plot (a, b) as well as the extracted diffusion exponent (c) and coefficient (d) for the Gauss-fit variance “fit" (orange triangles), profile variance “Var" (red squares), and width from the participation ratio PR (purple hexagons), and from IPW (green circles). 
    (a)~Ballistic expansion without disorder, $\disorderstrength = \SI{0}{\nano K} \times \boltzmann$. The black dashed line is Eq.~\eqref{eq:anomalous_diffusion} with fixed $\alpha = 2$ and $D_2 = E_\mathrm{F} / 8m$~\cite{giorgini_theory_2008}. Points plotted in lighter colors were omitted from the fitting procedure. Gauss-fit-variance error bars are standard errors from the fit, while errors of Var, PR, and IPW are calculated from error propagation.
    (b)~Expansion in weak disorder, $\disorderstrength = \SI{160}{\nano K} \times \boltzmann$, with point colors and symbols as in panel (a). Note that the vertical light-green line is the error bar of the final and omitted value of $\sigma_\mathrm{IPW}^2(t)$. The black dotted line is Eq.~\eqref{eq:anomalous_diffusion} with fixed $\alpha = 1$ and $D_1 = 10 \hbar / m$. This value of $D_1$ is not necessarily the expectation for this data set and is rather chosen to yield a guide to the eye, see Sec.~\ref{subsec:observable_performance}. 
    (c)~Diffusion exponent $\alpha$ as extracted from fits to the data shown in (a, data on top) and (b, data on the bottom). The expectation for the ballistic exponent $\alpha = 2$ (normal-diffusion exponent $\alpha = 1$) is plotted as a dashed (dotted) line. Areas shaded in gray show the results of a numerical investigation of observable performance (see Appendix~\ref{app:numerical_investigation} for details). Error bars are standard errors from fit. 
    (d)~Diffusion coefficient $D_\alpha$. The dashed line shows $D_2 = v_\mathrm{pr.}^2$. Error bars are standard deviations of values used to calculate $D_\alpha$ (see Eq.~\eqref{eq:diffusion_coefficient}) and are too small to be seen for most points. 
    }
    \label{fig3}
\end{figure*}

To choose the best observable for our system, we focus on the evaluation of two representative data sets, where the first data set corresponds to ballistic and the second to normal diffusion. 
We compare the extracted diffusion exponent $\alpha$ and generalized coefficient $D_\alpha$ from the respective observable with expectations for both data sets. 

For the first data set, see Fig.~\ref{fig3}(a), the disorder was absent ($\disorderstrength = 0$). 
Here, we predict a free (ballistic) expansion with $\alpha = 2$ and $D_2 = E_\mathrm{F} / 8m \eqqcolon v_\mathrm{pr.}^2$ (dashed line). 
The latter is the square velocity $v_\mathrm{pr.} \approx \SI{9.8}{\milli\meter\per\second}$ along the $y$ axis for a non-interacting Fermi gas released from a harmonic trap~\cite{giorgini_theory_2008}. 
As can be seen in Fig.~\ref{fig3}(a), the different observables are all generally in line with the expectation. 

For the second data set, shown in Fig.~\ref{fig3}(b), weak disorder $\disorderstrength = \SI{160}{\nano K} \times \boltzmann \approx E_\mathrm{F} / 4$ was applied to investigate diffusive expansion. 
Similarly to before, all observables generally follow the linear normal-diffusion behavior (dotted line) with $\alpha = 1$. 
Note that $\alpha = 1$ is only our expectation for this data set, as it is, to our knowledge, not possible to objectively infer \textit{a priori} which exact type of diffusion occurs at these weak disorder strengths. 
Still, we chose this data set because, for even smaller disorder strength, finite-size effects could effectively increase $\alpha$, as is described in more detail in Sec.~\ref{subsec:symmetric_disorder}. 
The diffusion coefficient $D_1 = 10 \hbar / m$, used for the dotted line in Fig.~\ref{fig3} (b), is set to yield a guide to the eye. 
As stated above, we aim to distinguish between normal diffusion and subdiffusion (for stronger disorder). 
Correspondingly, we consider the determination of $\alpha = 2$ as the more important benchmark for the observables. 

We show the performances of the observables in determining the diffusion exponent and coefficient, see Fig.~\ref{fig3}(c) and (d), respectively. 
Furthermore, for the exponent $\alpha$, we present the results of a numerical investigation indicated by gray areas. 
More specifically, we generated numerical data of a noisy Gaussian function whose width increased according to Eq.~\eqref{eq:anomalous_diffusion} either ballistically ($\alpha = 2$) or diffusively ($\alpha = 1$) and evaluated it identically to the measurement data. 
More details can be found in Appendix~\ref{app:numerical_investigation}. 
In the following, we focus on the performances of each observable separately. 

The Gauss-fit variance $\sigma^2_\mathrm{fit}$ (Sec.~\ref{subsec:gauss_fit_variance}) captures the ballistic expansion very well but claims subdiffusion for the second data set. 
For the ballistic case, the velocity $\sqrt{D_{2, \mathrm{fit}}} \approx \SI{9.7}{\milli\meter\per\second}$ we extract from the data set as described in Eq.~\eqref{eq:diffusion_coefficient} is the closest of all observables to the prediction $v_\mathrm{pr.} \approx \SI{8.9}{\milli\meter\per\second}$. 
Importantly, $\sigma^2_\mathrm{fit}$ appears to saturate toward larger $t$ in Fig.~\ref{fig3}(b), which could cause the lower exponent. 
A remaining curvature in a double-logarithmic plot signals a deviation from the modeled power-law behavior, meaning the analysis based on the anomalous diffusion as in Eq.~\eqref{eq:anomalous_diffusion} breaks down. 
As can be seen in Fig.~\ref{fig3}(c), the numerically determined Gauss-fit variance expectedly yields very accurate exponents as the underlying distribution function is precisely the same, which is experimentally not the case. 
As stated in Ref.~\cite{shapiro_cold_2012}, the density profile of fermions diffusing through disorder is expected to differ significantly from a Gaussian function. 
The profile would be modified even further when localization is considered. 
As we discuss in more detail in Appendix~\ref{app:density_profiles}, it is impossible to discern the underlying shape due to the large amount of image noise. 
Using another function, such as a generalized Gaussian, would technically fit better to the data (see Appendix~\ref{app:density_profiles}) but its parameters might not yield as much physical meaning while possibly suffering from the mentioned problem of possible overfitting. 

The exponent extracted from the variance of the density profile $\sigma^2_\mathrm{Var}$ (Sec.~\ref{subsec:distribution_variance}) significantly underestimates the ballistic exponent for the first data set but is compatible with normal diffusion in the second one. 
In both cases, most of its initial values have to be omitted due to them being negative. 
Note that negative values are physically not sensible but arise due to too much noise influence if the observable does not increase significantly enough for earlier expansion times and, thus, subtracting $\sigma_i^2(0)$ can result in negative values. 
This susceptibility to noise is also reflected in the large uncertainties in the numerical analysis.
The fact that $\sigma^2_\mathrm{Var}$ changes from strong fluctuations around the zero point to the power law of interest only for rather long expansion durations makes it problematic from the point of view of evaluation, as the signal of the data then approaches the magnitude of the camera noise (see Fig.~\ref{fig1}(c) for example), which makes its prediction less reliable.
We conclude that $\sigma^2_\mathrm{Var}$ is not a good observable for our system. 

The participation ratio (Sec.~\ref{subsec:participation_ratio}) has the obvious advantage that it is not dependent on external factors other than the atomic density detected. 
Moreover, it is quite robust against noise, at least compared to $\sigma^2_\mathrm{Var}$. 
According to the numerical investigation, $\sigma^2_\mathrm{PR}$ seems to be quite accurate in determining $\alpha$ for the case of normal diffusion, but it also finds a subdiffusive behavior for the experimental data. 
The major drawback, however, is the significant underestimation of the ballistic exponent for both the numerical and, especially, the experimental data set. 
Since it does not meet this benchmark, we cannot consider it a sufficiently good observable to study anomalous diffusion in our system accurately. 

The statistical observable $\sigma^2_\mathrm{IPW}$ (Sec.~\ref{subsec:ipw}) agrees very well with the expectations in both cases. 
The numerical data suggests that it tends to generally slightly overestimate both exponent and coefficient while being precise, especially in the case of normal diffusion. 
Indeed, the diffusion coefficient $D_1$ determined in the numerical investigation is consistently overestimated by a factor of less than two. 
For the coefficient determined for the ballistic case, we find $\sqrt{D_{2, \mathrm{IPW}}} \approx \SI{12.3}{\milli\meter\per\second}$ which is slightly larger than $v_\mathrm{pr.}$. 
Since IPW yields the largest estimation for the diffusion coefficient of the investigated observables, the experimentally determined coefficient can be seen as a close upper bound. 
An additional property of $\sigma^2_\mathrm{IPW}$ is that, for very large timescales, the variance and its error both suddenly increase significantly. 
This occurs when $w(t) \approx w_\mathrm{noise}$, causing $\sigma^2_\mathrm{IPW}$ to diverge. 
This has the advantage that the choice of the fit range is quite obvious, see $\sigma^2_\mathrm{IPW}$ and its error bar for the largest $t$ in Fig.~\ref{fig3}(b). 

We conclude from the comparison that both the Gaussian fit and IPW are good observables. 
However, extracting the diffusion exponent would be done from a fit of fit results for $\sigma^2_\mathrm{fit}$, since we perform a linear regression as in Eq.~\eqref{eq:log_anomalous_diffusion}. 
As the curvature of $\sigma^2_\mathrm{fit}$ is more strongly pronounced for expansions in stronger disorder, the precise choice of fit range would significantly influence the final exponent. 
In fact, this issue was the main motivation behind the search for alternative observables. 
Therefore, for the remainder of this work, we use the inverse participation width to determine the diffusion properties of the ultracold atom cloud.

\section{Signatures of Anderson Localization}
\label{sec:anderson_localization}

Using the inverse participation width, we analyze the behavior of the atoms when diffusing through disorder of different strengths and infer signatures of Anderson localization from deviations of normal diffusion and additional quantities. 
We first focus on the diffusion and localization fraction when the full cloud is symmetrically diffusing in a disordered potential in Sec.~\ref{subsec:symmetric_disorder}. 
Second, we directly compare the spatially resolved probability density distributions of clouds diffusing in asymmetric disorder in Sec.~\ref{subsec:asymmetric_disorder}. 
In this scenario, one part of the cloud diffuses into the disorder, while the other part shows close-to-ballistic expansion in very shallow disorder.

\subsection{Diffusion in symmetric disorder}
\label{subsec:symmetric_disorder}

\begin{figure*}
    \centering
    \includegraphics[width=178mm]{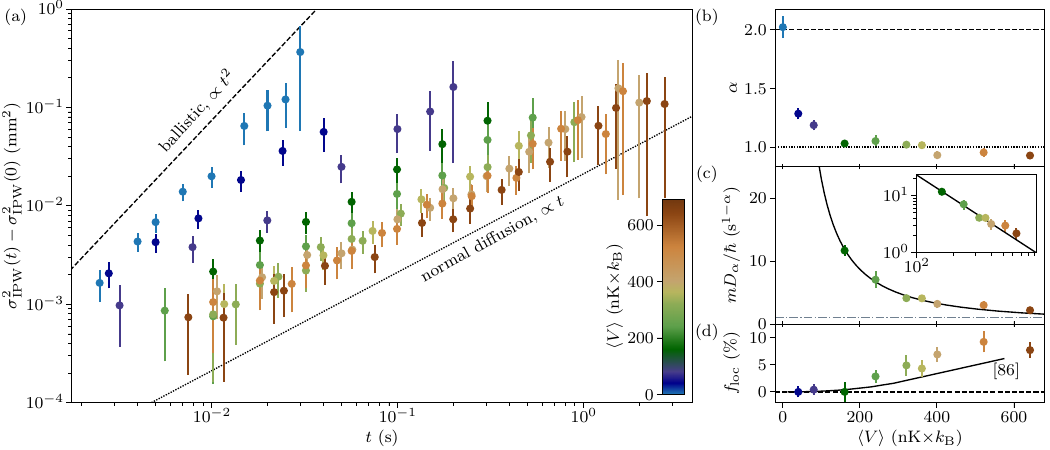}
    \caption{Expansion in disorder of variable strength. 
    Without disorder (light blue on the upper left side), the atoms expand ballistically. For very weak disorder (dark blue), $\disorderstrength < \SI{100}{\nano K} \times \boltzmann$, the system appears to be super-diffusive due to finite-size effects. For weak to moderate disorder (green), $\disorderstrength = \SIrange{100}{400}{\nano K} \times \boltzmann$, the atoms undergo normal diffusion. For strong disorder (brown), $\disorderstrength \geq \SI{400}{\nano K} \times \boltzmann$, subdiffusion is observed. 
    (a)~Atom-cloud variance from IPW over time for different disorder strengths (colors), plotted double-logarithmically. Error bars are calculated from error propagation. The power laws for ballistic expansion (dashed line) and normal diffusion (dotted line) are plotted as a visual aid. 
    (b)~Diffusion exponent $\alpha$ over disorder strength. The dashed (dotted) line marks the ballistic-expansion (normal-diffusion) exponent. Error bars are standard errors from fit.
    (c)~Diffusion coefficient $D_\alpha$ over disorder strength. Values of data with $\disorderstrength < \SI{100}{\nano K} \times \boltzmann$ were omitted to enhance visibility. The black line is a power-law fit which yields the exponent $-1.41 \, \pm \, 0.10$. The quantum of diffusion $\hbar / m$ is shown as a dash-dotted line. Error bars are standard deviations of values used to calculate $D_\alpha$. The inset shows the same data as a double-logarithmic plot to highlight the power-law behavior of $D_\alpha$. 
    (d)~Fraction of localized atoms $f_\mathrm{loc}$ over disorder strength. Error bars are standard errors from fit. The black solid line shows the theoretically expected localized fraction, see Eq.~\eqref{eq:floc_general}, estimated from the numerical results of Ref.~\cite{pasek_anderson_2017}. 
    }
    \label{fig4}
\end{figure*}

Experimentally, we align the focus of the speckle-laser beam to the cloud position to ensure that the cloud can symmetrically diffuse along the $y$ axis in the disorder potential. 
For each disorder strength we take a series of \textit{in-situ} absorption images for increasing diffusion time $t$ and extract the inverse participation width. 
An overview of the series of cloud variances for varying disorder strengths is shown in Fig.~\ref{fig4}(a). 
All expansion series follow the expected power-law behavior (straight line in a double-logarithmic plot). 
For disorder strengths $\disorderstrength > 0$, the system quickly transitions from ballistic expansion toward normal diffusion. 
There is an apparent intermediate regime of $1 < \alpha < 2$ for $\disorderstrength < \SI{100}{\nano K} \times \boltzmann$ (dark-blue points) which can be explained by finite-size effects. 
Faster atoms with energies close to the Fermi energy $E_\mathrm{F} \gg \disorderstrength$ perceive the disorder potential only as a small perturbation and can thus move approximately ballistically. 
Further, since the speckle-beam waist is finite at roughly \SI{450}{\micro\meter} along the $y$ axis, even the slower atoms can leave the central part of the disorder field during the diffusion time such that the speckle inhomogeneity has a substantial effect for these energies. 

The transition toward normal diffusion is also reflected by the exponent $\alpha$ shown in Fig.~\ref{fig4}(b).
For moderate disorder strengths, $\disorderstrength = \SIrange{100}{400}{\nano K} \times \boltzmann$ (green points), the diffusion exponent becomes constant at unity, as expected for normal diffusion. 
In that regime, the diffusion coefficient, shown in Fig.~\ref{fig4}(c), still decreases significantly for increasing disorder. 
In fact, we find that $D_\alpha$ also follows a power-law behavior (see black line) with a fit yielding an exponent of $\nu_D = -1.41 \, \pm \, 0.10$. 

For strong disorder, $\disorderstrength \geq \SI{400}{\nano K} \times \boltzmann$ (brown points), the exponent rather suddenly switches to that of slight but statistically significant subdiffusion, $\alpha = 0.93 \, \pm \, 0.03$. 
Deviations in the form of subdiffusion have been reported when interactions are introduced~\cite{derrico_quantum_2013}, however, these are entirely absent in this work due to Pauli blocking. 
Since subdiffusion with $\alpha = 2/3$ is expected to occur near the mobility edge~\cite{shapiro_cold_2012}, we interpret this observation as the emergence of localization effects hindering the diffusive expansion. 
The deviation of the observed exponent $2/3 < \alpha < 1$ may be attributed to the fact that the observable IPW is sensitive to the peak density. 
For a smaller fraction of particles starting to localize and the majority of particles undergoing normal diffusion, our observable accordingly shows a mixture of both behaviors. 
Especially taking into account that we release a degenerate Fermi gas with a relatively broad spectrum of initial energies and momenta, a large fraction of not localized, extended states will be present even for the highest disorder potentials.  
We note, however, that a direct comparison of the localization lengths in strong and weak disorder allows us to extract the predicted exponent of $\alpha = 2/3$ for weak localization, see Sec.~\ref{subsec:asymmetric_disorder}. 

The diffusion coefficient $D_\alpha$ lies well within the order of the quantum of diffusion $\hbar / m$~\cite{kuhn_coherent_2007, jendrzejewski_three-dimensional_2012, semeghini_measurement_2015, muller_comment_2014, shapiro_cold_2012, patel_universal_2020, sommer_universal_2011, enss_quantum_2012} for the largest disorder strengths. 
It is expected to vanish as $|\disorderstrength - V_\mathrm{c}|^{(d-2)\nu}$~\cite{lopez_experimental_2012, wegner_electrons_1976, abrahams_scaling_1979} with a critical disorder strength $V_\mathrm{c}$, the critical exponent $\nu \approx 1.58$~\cite{shapiro_cold_2012, lopez_experimental_2012, slevin_corrections_1999, MacKinnon1983}.
The dimensionality $d$ of our system is not straightforward to define. 
The expansion is limited to one dimension, while the atom cloud is always three-dimensional. 
Investigating a possible 1D-3D crossover will be an interesting topic for future works. 

While the exponent $\nu_D$ of $D_\alpha(\disorderstrength)$ seems to be compatible with the expectation from Anderson theory, it is unclear to what extent $\nu$ and $\nu_D$ can even be directly compared as no indication for some critical disorder strength is apparent in Fig.~\ref{fig4}(c). 
As mentioned above, the competing energy scales of our system make it difficult to draw a reliable conclusion. 
We are also not sure whether we should expect $D_\alpha$ to vanish at all since the largest contribution to the observation is always diffusive.

To support our interpretation, we first employ the widely used Ioffe-Regel criterion to estimate whether our system should be expected to exhibit localization effects in principle. 
It compares a particle's mean free path (often approximated by the disorder's correlation length $\eta$, see below) with its wavelength, usually the DeBroglie wavelength $\lambda_\mathrm{dB} = \hbar \sqrt{2\pi / m k_\mathrm{B} T}$~\cite{kondov_disorder-induced_2015}. 
Our setup fulfills this criterion. 

However, as shown in Refs.~\cite{delande_mobility_2014, pasek_anderson_2017}, the actual mobility edge is significantly overestimated by this approach. 
Hence, we instead compute the critical momentum $k_\mathrm{AL}$ below which AL can occur according to Ref.~\cite{beilin_diffusion_2010}. 
This critical momentum is explicitly adapted to the case of small correlation energies $E_\mathrm{\eta} = \hbar^2 / (m \eta^2)$ of the disorder. 
More specifically, our experiment is in the regime of $\disorderstrength / E_\mathrm{\eta} > 1$ if $\disorderstrength > \SI{25}{\nano K} \times \boltzmann$ for the geometric mean of correlation lengths $\overline{\eta}$ (and $\disorderstrength > \SI{150}{\nano K} \times \boltzmann$ for the transversal correlation lengths $\eta_{x,y}$). 
For that case specifically, we compute the critical momentum as $k_\mathrm{AL} \approx (\disorderstrength / E_\mathrm{\eta})^{2/5} / \eta$~\cite{beilin_diffusion_2010}. 
Comparing it with the Fermi momentum $k_\mathrm{F}$ for the strongest disorder yields $k_\mathrm{F} / k_\mathrm{AL} \approx 1.8$ for $\eta = \overline{\eta}$ (and $k_\mathrm{F} / k_\mathrm{AL} \approx 0.8$ for $\eta = \eta_{x,y}$). 
Therefore, we expect at least a significant fraction of the fermions with smaller momenta to localize. 

To investigate this onset of localization quantitatively, we adopt the method to infer the localized fraction $f_\mathrm{loc}$ as introduced in Ref.~\cite{jendrzejewski_three-dimensional_2012}. 
This value estimates the infinite-time fraction of atoms that would not diffuse away due to localization, assuming no losses. 
We modify the method for our expansion along only the $y$ axis and implement the complete anomalous-diffusion power law from Eq.~\eqref{eq:anomalous_diffusion} into the fit function. 
We use the histogram width to approximate the peak density as introduced in Sec.~\ref{subsec:ipw}. 
Thus $f_\mathrm{loc}$ is the only free parameter. 
See Appendix~\ref{app:floc} for more details on its evaluation. 

The localized fraction is shown in Fig.~\ref{fig4}(d), and it is expectedly zero for low disorder strengths. 
It starts to increase only around the disorder strength $\disorderstrength \approx \SI{250}{\nano K} \times \boltzmann$. 
The disorder strength where the localized fraction starts to form roughly coincides with the disorder range where the diffusion coefficient enters the order of magnitude of the diffusion quantum $\hbar / m$. 
The largest value we observe is $f_\mathrm{loc} = \SI{9.2 \pm 2.0}{\%}$. 
This agrees well with the hypothesis that diffusion is the main contribution to our observation, reflecting influences of Anderson localization when the disorder becomes sufficiently strong. 
The course and order of magnitude of $f_\mathrm{loc}$ as well as the general behavior of the diffusion coefficient agree well with the findings of Ref.~\cite{jendrzejewski_three-dimensional_2012}. 

Finally, we compare the extracted localized fraction to the theoretical expectation from Ref.~\cite{pasek_anderson_2017}, where the mobility edge $E_\mathrm{c}$ was investigated numerically for anisotropic 3D systems. 
We write the localized fraction for the case $E_\mathrm{F} > E_\mathrm{c}$ more generally as
\begin{equation}
\label{eq:floc_general}
    f_\mathrm{loc} = \frac{\int_{0}^{E_\mathrm{c}} \rho(E) \mathrm{d}E}{\int_{0}^{E_\mathrm{F}} \rho(E) \mathrm{d}E},
\end{equation}
where $\rho(E)$ is the density of states. 
Since our gas is degenerate and initially prepared in a three-dimensional harmonic trap, all states with energies up to $E_\mathrm{F}$ are populated according to the density of states $\rho(E) \sim E^2$~\cite{giorgini_theory_2008}. 
For this evaluation, we employ the naive assumption that this density of states is still valid when the disorder is introduced. 
A more accurate evaluation of the density of states by a numerical analysis according to Ref.~\cite{jendrzejewski_three-dimensional_2012} is beyond the scope of this paper and will be left for future work. 
Thus, we compute $f_\mathrm{loc} = E_\mathrm{c}^3 / E_\mathrm{F}^3$. 
The solid line in Fig.~\ref{fig4}(d) shows the theoretically expected localized fraction as calculated from this energy ratio using the numerical results from Ref.~\cite{pasek_anderson_2017}. 
Here, we used the transversal correlation lengths for the correlation energy. 
Our experimental data systematically overestimates $f_\mathrm{loc}$ slightly but generally agrees with the expectation. 
The larger fraction could be an effect of the presumed 1D-3D crossover since theoretical works that investigate strongly anisotropic settings suggest that anisotropy reduces the dimensionality and enhances localization effects~\cite{zhang_anderson_1990, chu_anderson_1993}.

\subsection{Diffusion in asymmetric disorder}
\label{subsec:asymmetric_disorder}

\begin{figure*}
    \centering
    \includegraphics[width=178mm]{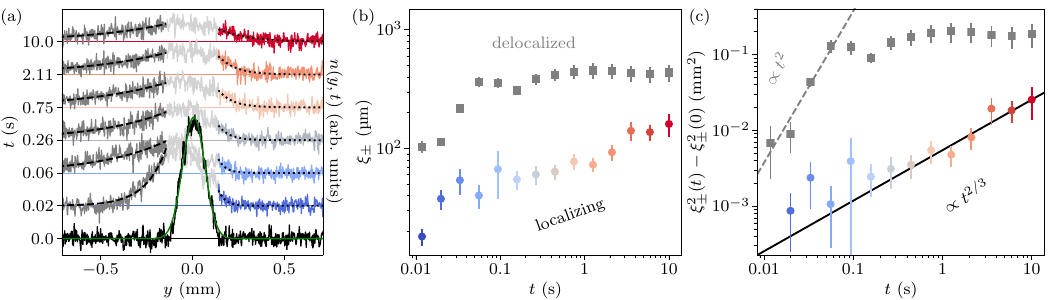}
    \caption{Asymmetric-disorder measurement for variable expansion time and fixed $\disorderstrength = \SI{691}{\nano K} \times \boltzmann$. 
    (a)~Density profiles $n$ for increasing expansion time $t$ (bottom to top, left axis). The colored lines at the right side (dark-gray lines at the left side) denote the part in $+y$ ($-y$) direction outside the extent of the cloud at $t = 0$, which itself is plotted as light-gray lines around $y = 0$. The outer regions are fitted with exponential functions (left: black dashed lines, right: black dotted lines)
    The bottom-most line plot (black line) shows the $t=0$ density profile and a Gaussian fit (green dashed line). 
    The density scale (right axis) is the same for all profiles. The horizontal solid lines mark $n(y, t)=0$ for each respective $t$ in the same color. 
    (b)~Cloud-extension length scale $\xi_\pm$ as extracted from density-profile fits over time. Gray squares show values for $-y$ direction, away from the disorder, and colored circles for $+y$ direction, toward it. Error bars are standard errors from fit.
    (c)~Variance $\xi^2_\pm(t) - \xi^2_\pm(0)$ over time with the same point colors and symbols as in panel (b). The dashed gray line is a power-law fit with fixed exponent $\alpha = 2$ to the first four points of $-y$ direction. The black line is a power-law fit with fixed exponent $\alpha = 2/3$ to all plotted $+y$-direction data. 
    }
    \label{fig5}
\end{figure*}

Beyond the cloud width, an additional quantity often used to characterize localization is the localization length~$\xi$, i.e., the length scale on which a localized wave function decays. 
For the non-equilibrium diffusion considered here, the signatures of localization on the atomic density distribution are mixed with signatures from diffusion.
To unravel the contributions of disorder-induced localization and diffusion, we modify the setup by displacing the speckle-disorder beam toward the $+y$ direction. 
In this setting, the Fermi gas is initially trapped at the edge of the disorder envelope, and the atom cloud released experiences a strong disorder in the $+y$ direction and a weak or even negligible disorder in the $-y$ direction. 
The resulting time-resolved density distributions for the maximally achievable disorder strength of $\disorderstrength \approx \SI{691}{\nano K} \times \boltzmann$ is shown in Fig.~\ref{fig5}. 
While the resulting density distribution in the $-y$ direction can be assumed to be free of any localization effects, the part in the $+y$ direction will be affected by localization and diffusion. 
This version of the setup is reminiscent of transfer-matrix approaches to probe localization in the sense that the transport of particles moving toward the disorder is limited by the probability of transmission versus reflection. 

During a few tens of milliseconds, the cloud shape changes from the trapped-gas profile to a distribution that is compatible with exponential functions on both tails outside of its extension at $t=0$ (see first three line plots from bottom to top in Fig.~\ref{fig5}(a)). 
The low signal-to-noise ratio does not allow us to distinguish between exponential, stretched exponential, and power-law functions (see Appendix~\ref{app:density_profiles}). 
Still, to investigate the behavior of the length scale, we fit an exponential function (see dotted or dashed lines), which yields a cloud-extension length scale $\xi_\pm$ along the $\pm y$ direction. 

After roughly \SI{60}{\milli\second}, $\xi_-$ (gray squares) saturates to an equilibrium value, while the other side continues to increase, albeit slowly, see Fig.~\ref{fig5}(b). 
In fact, the variance $\xi^2_-(t) - \xi^2_-(0)$ reveals a ballistic expansion in the direction away from the disorder (gray squares) before saturating. 
Conversely, the cloud moving toward the disorder expands subdiffusively with exponent $\alpha = 2/3$. 
We confirm this by fitting power laws onto the variances where the only free parameter is the diffusion coefficient, see lines Fig.~\ref{fig5}(c). 

The velocity from the ballistic diffusion coefficient is roughly $\sqrt{D_2} = \avg{v_-} \approx \SI{6}{\milli\meter\per\second}$ which is somewhat lower than the velocity found in the disorder-free expansion, see Sec.~\ref{subsec:observable_performance}. 
This confirms that some amount of disorder is still present in that region, explaining the significantly prolonged observation duration achieved compared to the data set shown in Fig.~\ref{fig4}. 
For the subdiffusion in the strong-disorder region, we find $m D_{2/3} / \hbar = \SI{0.26 \pm 0.03}{\second}^{1/3}$. 
Since the exponent of $\alpha = 2/3$~\cite{shapiro_cold_2012} agrees very well with the data and the fitted diffusion coefficient lies below the quantum of diffusion~\cite{kuhn_coherent_2007, jendrzejewski_three-dimensional_2012, semeghini_measurement_2015, muller_comment_2014, shapiro_cold_2012, patel_universal_2020, sommer_universal_2011, enss_quantum_2012}, we conclude that the system must be close to or at the mobility edge as these observations are clear signatures for the Anderson transition.

\begin{figure*}
    \centering
    \includegraphics[width=178mm]{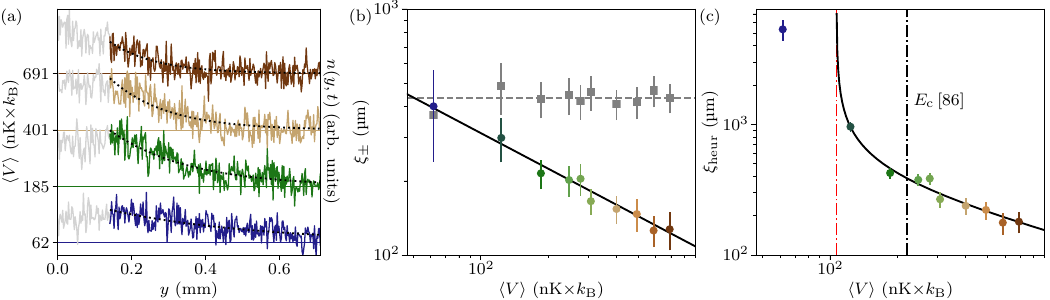}
    \caption{Asymmetric-disorder measurement for variable disorder strength and fixed $t = \SI{1}{\second}$. 
    (a)~Density profiles $n$ for increasing disorder strength \disorderstrength{} (bottom to top, left axis). The colored line denotes the part outside the extent of the cloud at $t = 0$ (gray lines) and marks the data fitted by an exponential function (black dotted lines). 
    The density scale (right axis) is the same for all profiles. The horizontal solid lines mark $n(y, t)=0$ for each respective \disorderstrength{} in the same color. 
    (b)~Cloud-extension length scale $\xi_-$ (gray squares) and $\xi_+$ (colored circles) direction as extracted from density-profile fits. The black line is a power-law fit of $\xi_+$ with exponent $-0.48\, \pm \,0.01$. The gray dashed horizontal line shows $\avg{\xi_-}$. Error bars are standard errors from fit.
    (c)~Heuristic length scale for localization $\xi_\mathrm{heur}$. The black line is a power-law fit of type $|\disorderstrength - V_\mathrm{c}|^{-\nu}$ which yields $\nu = 0.47\, \pm \,0.09$ and $V_\mathrm{c} = \SI{107 \pm 9}{\nano K} \times \boltzmann$ (red dash-dotted line). 
    Error bars are calculated from error propagation. 
    The black dot-dashed line shows the mobility edge $E_\mathrm{c}$ as estimated from the results of Ref.~\cite{pasek_anderson_2017} for our disorder. 
    }
    \label{fig6}
\end{figure*}

The $+y$ extensions of the density profiles after a fixed expansion time of $t = \SI{1}{\second}$ for increasing disorder strengths are shown in Fig.~\ref{fig6}. 
We chose this duration as a compromise between ensuring a sufficiently long expansion and, hence, clear signatures on the one hand and avoiding losses from collisions with background particles on the other hand. 
The cloud-extension length scale $\xi_+$ appears to follow a power-law behavior (black line in Fig.~\ref{fig6}(b)) with an exponent of $-0.48\, \pm \,0.01$. 
In this direct plot of the length scale $\xi_+$, no critical behavior as expected from Anderson theory can be observed. 
We emphasize, however, that the effect of extended states also influences the observation in strong disorder. 
To compensate for the contribution of extended states, we introduce a heuristic localization length 
\begin{equation}
    \xi_\mathrm{heur} = \frac{\xi_+}{1 - \xi_+ / \avg{\xi_-}}. 
\end{equation}
We motivate this length scale as a rescaling of $\xi_+$, which we expect to exhibit signatures of localized states, with the diffusion-dominated length scale $\xi_-$, which we know to be delocalized. 
For the latter, we use the average value $\avg{\xi_-} = \SI{435 \pm 31}{\micro\meter}$ since $\xi_-$ does not depend on the disorder strength $\disorderstrength$. 
In the limiting case  $\xi_+ \ll \xi_-$, the length scale $\xi_\mathrm{heur}$ is equal to $\xi_+$, indicating that $\xi_+$ is close to the localization length, while $\xi_\mathrm{heur} \rightarrow \infty$ for $\xi_+ = \xi_-$, and the system behaves as in the disorder-free case. 

In Fig.~\ref{fig6}(c), $\xi_\mathrm{heur}$ is shown with a power-law fit of type $A |\disorderstrength - V_\mathrm{c}|^{-\nu}$, containing a critical disorder strength $V_\mathrm{c}$, the critical exponent $\nu$ and a prefactor $A$ as free parameters. 
This function agrees reasonably well with the coarse of $\xi_\mathrm{heur}$ and yields an exponent $\nu = 0.47 \, \pm \, 0.09$ and $V_\mathrm{c} = \SI{107 \pm 9}{\nano K} \times \boltzmann$. 
A critical exponent of $0.5$ has been reported in the literature~\cite{aegerter_experimental_2006, schuster_relation_1978}. 
However, the generally accepted critical exponent for the Anderson transition is $\nu \approx \SI{1.58}{}$~\cite{shapiro_cold_2012, lopez_experimental_2012, slevin_corrections_1999}. 
The exponent we find disagrees with this value. 
From the numerical result of Ref.~\cite{pasek_anderson_2017}, we estimate a mobility edge of $E_\mathrm{c} \approx \SI{220}{\nano K} \times \boltzmann$ for our system. 
We compute this value using the transversal correlation length $\eta_{x, y}$ for the correlation energy $E_\eta$ and the theoretical mobility edge at $\disorderstrength / E_\eta = 4$, which matches the settings with stronger disorders in our experiment. 
Thus, we find a critical disorder strength that lies below $E_\mathrm{c}$ by a factor of roughly two. 
However, as $\xi_\mathrm{heur}$ is a purely heuristic length scale, it is unclear if it is at all expected to exhibit the Anderson critical power law. 
As mentioned above, we suspect that our system might be on a 1D-3D crossover. 
Correspondingly, neither pure-1D nor pure-3D Anderson theory could be fitting expectations. 
On the one hand, the anisotropy in this setting is not expected to change the critical exponents \cite{Milde2000, Goichoechea2021Aop}. 
On the other hand and as mentioned above, theoretical works suggest enhanced localization in strong anisotropy~\cite{zhang_anderson_1990, chu_anderson_1993}.
To our knowledge, it is so far unexplored how such a crossover influences the critical exponent. 
We emphasize that this analysis is not supposed to determine the critical exponent but, contrarily, suggests that focusing exclusively on exponential density profiles is insufficient when investigating Anderson localization.

\section{Conclusion and Outlook}
In summary, we experimentally investigated the competition of diffusion and localization in an ultracold non-interacting Fermi gas with relatively broad initial energy and momentum distributions. 
We presented and compared four observables for MSD of \textit{in-situ} images of the ultracold atom cloud, which can be used to investigate diffusion and localization. 
We further carefully examined how our system crosses over from being compatible with pure normal diffusion to a subdiffusion, being influenced by Anderson localization. 
In a displaced-disorder configuration, we could observe the power law of $t^{2/3}$ that is expected near the mobility edge before the system becomes fully localized~\cite{shapiro_cold_2012}. 
We further emphasize that the observation of density distributions which can be described by exponential functions is not sufficient to identify Anderson localization unequivocally as exponential tails would even be expected in a purely diffusive setting~\cite{shapiro_expansion_2007, shapiro_cold_2012}. 

A thorough investigation of the dimensionality of our setup will be of direct interest as we could map out the supposed 1D-3D crossover by changing the radial trapping frequencies. 
As mentioned above, the extension to smaller-grained and more isotropic disorder should further allow for a thorough investigation of the mobility edge. 
Since our system is generally capable of creating a strongly interacting gas of both bosonic and fermionic nature~\cite{ganger_versatile_2018, koch_quantum_2023}, we could investigate the influence of quantum statistics and thus initial energy distribution, inter-particle interactions, and even superfluidity. 
In recent years, machine learning (ML) concepts for data analysis in physical systems have shown widespread use and better performances in detecting anomalous diffusion than other common techniques \cite{munoz-gil_objective_2021}.
Combining this with possible ML regression models, one may extract the physical features of the noisy density profiles in interrelation with the newly presented density histograms. Furthermore, statistical methods for deconvolutions are mostly not robust, and their quality depends strongly on the strength of the noise \cite{Carroll1988}. 
Here, ML non-linear and generative models also promise more robust approaches \cite{Zenil2019, Kingma2021}. 
Sampling techniques and data augmentation can be equally useful for image and signal processing, particularly when augmentation coincides with the effects expected in the underlying physical system \cite{palaiodimopoulos2023quantum}.

\section*{Data availability statement}
All data of the figures in the manuscript and Methods are available in a Zenodo repository: \href{https://zenodo.org/records/10697491}{https://zenodo.org/records/10697491} (Ref.~\cite{zenodo_static}). 

\section*{Acknowledgements}
We thank M. Fleischhauer, R. Unanyan, A. Go\"icoechea, V. Fortes Rey, D. Hern\'andez-Rajkov, G. Roati, B. Shapiro, and G. Orso for fruitful discussions as well as M. Kaiser and A. Guthmann for carefully reading the manuscript. 
Additional thanks to G. Orso for providing the numerical result from his work in Ref.~\cite{pasek_anderson_2017}. 
This work was supported by the German Research Foundation (DFG) by means of the Collaborative Research Center Sonderforschungsbereich SFB/TR185 (Project 277625399). M.K.-E. acknowledges support by the Quantum Initiative Rhineland-Palatinate QUIP.
J.K. acknowledges support by the Max Planck Graduate Center with the Johannes Gutenberg-Universität Mainz.

\appendix

\section{Details on experimental methods}
\label{app:methods}

\begin{figure}
    \centering
    \includegraphics[width=86mm]{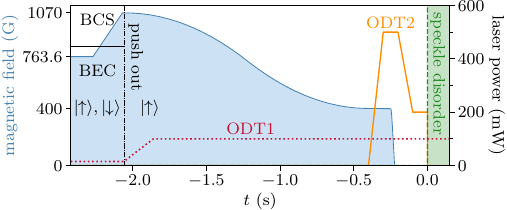}
    \caption{Sketch of the preparation of the spin-polarized gas. 
    The magnetic field (large blue area) pertains to the left axis, and the laser powers (ODT1\&2, speckle disorder) to the right axis. 
    The black horizontal line shows the Feshbach resonance for the spin mixture in states $\ket{\uparrow}, \ket{\downarrow}$ at $B = \SI{832.2}{G}$. 
    The time of the push-out beam is shown as a black dash-dotted line. 
    }
    \label{fig7}
\end{figure}

As stated in the conclusion, our system can create strongly interacting gases. 
More specifically, we usually work with an equal mixture of the two spin states $\ket{\uparrow} \coloneqq \ket{m_J = -1/2, m_I=1}$ and $\ket{\downarrow} \coloneqq \ket{m_J = -1/2, m_I=0}$ at high magnetic fields, with $m_J$ ($m_I$) being the magnetic quantum number of the electronic (nuclear) spin~\cite{nagler_cloud_2020}. 
We can apply magnetic fields of up to $B = \SI{1070}{G}$, granting us access to the broad Feshbach resonance around $B = \SI{832.2}{G}$ and, therefore, the crossover between Bose-Einstein-condensate (BEC) and Bardeen-Cooper-Schrieffer-type (BCS) superfluidity~\cite{grimm_ultracold_2007, zurn_precise_2013}. 
For more details about our setup, see Refs.~\cite{ganger_versatile_2018, nagler_cloud_2020}. 

In Fig.~\ref{fig7}, a sketch of the preparation of the spin-polarized gas is shown. 
We start the sequence used for this work by creating a BEC at $B = \SI{763.6}{G}$, followed by a slow ramp of \SI{200}{\milli\second} to $B = \SI{1070}{G}$. 
There, deep in the BCS regime where the fermionic pairs are weakly bound and spatially far apart, we apply a push-out laser pulse resonant to state $\ket{\downarrow}$ with a duration of \SI{2}{\micro\second}. 
During that pulse, only roughly \SI{10}{\%} of atoms in state $\ket{\uparrow}$ are lost due to resonant scattering, and no measurable amount of $\ket{\downarrow}$ atoms remain. 

Afterward, we deepen the trap by increasing the laser power of ODT1 from the initial \SI{15}{\milli\watt} to \SI{100}{\milli\watt} and perform a very slow (\SI{1.65}{\second}) double-parabolic field ramp down to an intermediate field of $B = \SI{400}{G}$. 
This is necessary because the position of our magnetic field center changes significantly with $B$, making this a transport over a distance of more than \SI{300}{\micro\meter}. 
Since our non-interacting sample cannot thermalize, any excitations will remain in the gas, which is why this ramp was chosen with such a long timescale. 
For every step of the sequence until $t=0$, we ensured no oscillations or unaccounted broadening occurred. 
Further, we never observe atom losses, the only exception being during the push-out pulse. 

At that stage, we load our gas into the crossed trap by slowly introducing ODT2 over \SI{100}{\milli\second} until a power of \SI{500}{\milli\watt} is reached. 
Then, we switch off the magnetic field rather quickly during \SI{30}{\milli\second}, also in a double-parabolic ramp. 
We found that switching off the field rapidly rather than slowly induced no measurable excitations. 
We explain that with the magnetic trap being relatively shallow at that stage ($\omega_y < 15 \times 2\pi\,$Hz), while ODT2's influence was large during the field shut off (resulting in $\omega_y > 60 \times 2\pi\,$Hz and several hundred $2\pi\,$Hz for the remaining axes). 

Once the magnetic field is switched off, the signal-to-noise ratio is significantly reduced due to optical pumping into dark states as described in Refs.~\cite{kinast_phd, gehm_phd}. 
Further, our imaging is aligned and calibrated for high magnetic fields, both of which explain the large noise in the images. 

To determine the atom number accurately, we ran an additional sequence in which we took absorption images after every step and after additionally reverting to the previous step. 
By reverting, we ensured that no atoms were actually lost even if the measured atom number was significantly different at $B = \SI{0}{G}$ compared to finite fields. 
From that measurement, we found an imaging-correction factor of roughly $3.4$, which translated the perceived atom number at zero field to the value we would measure at the field of $B = \SI{763.6}{G}$ for which our imaging is calibrated. 

Once the field has been switched off, we ramp the laser power of ODT2 down to \SI{200}{\milli\watt} during \SI{100}{\milli\second}. 
This finishes the preparation, after which we initialize the expansion at $t=0$ by switching off ODT2 while switching on the disorder. 
Using acousto-optical modulators, this step happens in less than one microsecond, faster than the timescale of the atom's motion. 
Finally, by instead imaging the cloud \textit{in situ}, we can extract its temperature as described in Refs.~\cite{hadzibabic_twospecies_2002, kinast_phd}.

\section{Statistical investigation of density distributions}
\label{app:histograms}

For the theoretical investigation of the qualitative profiles of $p(n,t)$ in the main text, we evaluate the normalized histograms for the continuum
\begin{align}
\label{eq:histograms}
    \wp(n,t)= \frac{1}{N(t)} \int_{-\infty}^{\infty} \mathrm{d}y \, \delta(n-n(y,y_0,t)),
\end{align}
where $\int_{w(t)} n \wp(n,t) \, \mathrm{d}n = 1$ which is given by the condition $\int_{-\infty}^{\infty} n(y,y_0,t) = N(t)$. 
For the remainder of this section, we will fix the peak position to $y_0 = 0$. 
Furthermore, $\wp(n,t)$ can be thought of as a kernel of a probability density function, as $\wp(n,t)$ is expected to show divergences in the continuum leading to $\int_{w(t)} \wp(n,t) \,\mathrm{d}n$ being not convergent. 
As a kernel, it has all the properties of a PDF besides normalization. 
Still, evaluating Eq.~\eqref{eq:histograms} allows for an easier analytical insight, which will become relevant when we test the control distribution, see Eq.~\eqref{eq:gauss_wavefunction}. 
Given an initial unimodal distribution in space without noise, see Eq.~\eqref{eq:gauss_wavefunction}, the normalized histogram has the form 
\begin{align}
    \wp_\mathrm{Gauss}(n,t) = \frac{\sqrt{2} \sigma}{n\sqrt{-\mathrm{ln}(\frac{n \sqrt{2 \pi \sigma^2(t)}}{N(t)})}},
    \label{eq:histgauss}
\end{align}
for the support $0 < n < \frac{N(t)}{\sqrt{2 \pi \sigma^2(t)}}$.
The two divergencies correspond to the most common occurrences of densities, which are 0 and the peak density. 
In the noise-free case, the peak density is then trivially given by $n(0,t) = \frac{N(t)}{\sqrt{2\pi\sigma^2(t)}}$ for the density centered around zero.

To further estimate the range of unimodal distributions in space, giving divergent flanks in their corresponding histogram, we assume a generalized Gaussian distribution centered around zero
\begin{equation}
\label{eq:generalized gauss}
   n_\mathrm{gG}(y, t)= \frac{\nu_\mathrm{gG}}{2 \sqrt{2 \sigma^2(t)} \, \Gamma(1/\nu_\mathrm{gG})} \mathrm{e}^{-\left(\frac{|y|}{\sqrt{2 \sigma^2(t)}}\right)^{\nu_\mathrm{gG}}},
\end{equation}
where $\Gamma(\cdot)$ is the gamma function and $\nu_\mathrm{gG}$ is an additional parameter controlling the shape of peak and tails. 
For $\nu_\mathrm{gG} = 2$, we restore the normal distribution and, for $\nu_\mathrm{gG} = 1$, we get a symmetric exponential distribution, as expected in the perfectly localized case~\cite{laflorencie_entanglement_2022}. 
Evaluating the histogram of Eq.~\eqref{eq:generalized gauss}, we get
\begin{equation}
    \wp_\mathrm{gG}(n, t, \nu_\mathrm{gG})=\frac{2 \sqrt{2 \sigma^2(t)}}{n \nu_\mathrm{gG} \left[ -\ln{\frac{n \sqrt{2 \sigma^2(t)} \Gamma(1 / \nu_\mathrm{gG})}{N(t) \nu_\mathrm{gG}}}\right]^{1 - 1 / \nu_\mathrm{gG}}},
\end{equation}
for the support $0 < n < \frac{N(t) \nu_\mathrm{gG}}{\sqrt{2 \sigma^2(t)} \Gamma(1 / \nu_\mathrm{gG})}$ which restores the $\wp_\mathrm{Gauss}$ for $\nu_\mathrm{gG} = 2$. This distribution has divergent flanks for all $\nu_\mathrm{gG} > 1$.
The divergent flanks in the histogram are necessary for the evaluation of IPW with noise since the convolution with the noise shows a bimodal profile from which we can approximate the noise-free width and, therefore, the peak density, see Sec.~\ref{subsec:ipw}. 

\begin{figure}
    \centering
    \includegraphics[width=86mm]{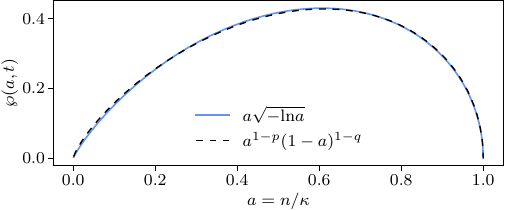}
    \caption{Comparison of inverse kernels of $\wp_\mathrm{Gauss}$ (blue solid line) and $\mathrm{Beta}$ (black dashed line), Eq.~\eqref{eq:beta_function}, over $a = n / \kappa$. For $p = 0.22$ and $q = 0.505$, as used here, the Beta distribution approximates $\wp_\mathrm{Gauss}$.
    }
    \label{fig8}
\end{figure}

Instead of assuming Gaussian density distributions in space, we may model the histogram. 
In statistics, it is common to model probabilities or random variables in a finite range $(0, n_\mathrm{max})$ via a Beta distribution~\cite{gupta2004handbook}  
\begin{align}
\label{eq:beta_function}
    \mathrm{Beta}(p, q, a) =\frac{a^{p - 1}\left(1 - a\right)^{q - 1}}{B(p, q)}, \\
    \mathrm{for} \ 0 < a < 1 \quad \mathrm{and} \quad B(p, q)=\frac{\Gamma(p) \Gamma(q)}{\Gamma(p + q)}.
\end{align}
where $p$ and $q$ are parameters and depend on the provided density profile. 
To fit the profile, we will look at the stretched Beta distribution $\kappa \mathrm{Beta}(p, q, \kappa a)$, where $\kappa = n(0, t) = w(t)$ is the stretching factor corresponding to the width $w(t)$ and $n = \kappa a$. 
If the density profile $n(y, t)$ is unknown, we can either fit or use the method of moments to evaluate $p$ and $q$. 
If $n(y, t)$ is known, we can fix $p$ and $q$, as shown in Fig.~\ref{fig8} for the exemplary case of a Gaussian, Eq.~\eqref{eq:histgauss}. 
We may also evaluate the variance of a variable $A \sim \mathrm{Beta}$ following the Beta distribution 
\begin{equation}
    \mathrm{Var(A)} = \frac{p q}{(p + q)^2(p + q + 1)}.
\end{equation}
Evaluating the variance for the stretched Beta distribution, where $p$ and $q$ are assumed constant, we receive $\kappa^2\mathrm{Var(A)}$. 
This means that the change of the variance is completely determined by the stretching factor, which is equal to the width $w(t)$.

\section{Numerical investigation of observable performance}
\label{app:numerical_investigation}

\begin{figure}
    \centering
    \includegraphics[width=86mm]{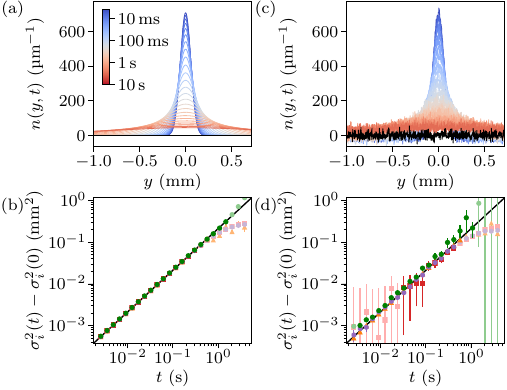}
    \caption{Observable performance on numerically generated data for the (a,~b) noise-free and (c,~d) noisy case. 
    (a,~c)~Generated Gaussian line densities over $y$ for different expansion times $t$, see colorbar in (a). 
    (b,~d)~Extracted $\sigma^2_i(t) - \sigma^2_i(0)$ over time $t$. The observables $i$ stand for the Gauss-fit variance “fit" (orange triangles), density-profile variance “Var" (red squares), width from the participation ratio PR (purple hexagons), and IPW (green circles), with same point colors and symbols as in Fig.~\ref{fig3}. 
    The black line shows the true $\sigma^2$ set for the plotted Gaussians. For the data shown here, $\alpha = 1$ and $D_1 = 10 \hbar / m$ is used. Errors are computed as for the experimental data. See Sec.~\ref{subsec:observable_performance} and Fig.~\ref{fig3} for details about the observables. 
    }
    \label{fig9}
\end{figure}

To further analyze the observable performances, we evaluated numerically generated density profiles. 
We begin by generating a Gaussian density, see Eq.~\eqref{eq:gauss_wavefunction}, over $y$ from a Gauss fit to the trapped-gas profile from the experimental data. 
We use a $y$ axis analogous to the axis in the experiment to simulate our imaging system's resolution and image size. 
For a given list of expansion times $t$, we generate new Gaussians for every $t$ while changing their $\sigma(t)$ according to Eq.~\eqref{eq:anomalous_diffusion}, simulating a diffusive expansion. 
Depending on the case, we use $\alpha = 1$ or $2$ and any arbitrary $D_\alpha$. 
Before we evaluate the line densities in the same way as for the experimental data, we can optionally add random white (Gaussian) noise, which we usually set to have the same $\sigma_\mathrm{noise}$ as the camera noise, see Fig.~\ref{fig2} and Sec.~\ref{subsec:ipw}. 
As in the experiment, we generate 50 images per setting that are averaged and then evaluated further. 
In Fig.~\ref{fig9}, the noise-free density distributions are shown in (a), and the data with noise in (b). 
The various $\sigma^2_i$ for the respective cases are plotted in (c) and (d). 

Compared to the experimental data, this investigation has several advantages. 
First, we can switch off the “camera" noise to investigate its influence on the results directly. 
Similarly, it is significantly faster to generate more data numerically compared to running the experiment, yielding better statistics. 
Second, we can increase or decrease both resolution and “image" size, which allows us to eliminate or enhance finite-size effects. 
Further, since we define how the drawn densities expand in time, we have a true reference to compare the inferred results to (see solid line in Fig.~\ref{fig9}(c,~d)). 
Also, we are not limited by long-time atom losses. 

Note that this investigation effectively only yields information about how well a perfect (albeit noisy) Gaussian can be evaluated during an expansion similar to that of our experiment. 
As we interpret our cloud as being partially localized and, therefore, potentially bimodal, this is not captured by this effort. 
It would be, however, interesting how different cloud shapes (see, e.g., the profiles shown in Fig.~\ref{fig10}) would be evaluated. 
Nevertheless, we can infer useful qualitative information about the observable performance, i.e., that $\sigma^2_\mathrm{Var}$ appears very susceptible to noise in general. 
Further, independent of noise but enhanced by it, all observables tend to curve below the simulated $\sigma^2$, except for IPW, which is observed to do the opposite. 
Nevertheless, with sensible fit-range choices, all observables perform perfectly in determining both diffusion quantities if no noise is present. 
For the results in the case with noise, see Fig.~\ref{fig3}(c). 
Overall, combining such a supporting numerical investigation with the evaluation of the experimental data yields a good overview of the observable performances.

\section{Density profiles}
\label{app:density_profiles}

\begin{figure}
    \centering
    \includegraphics[width=86mm]{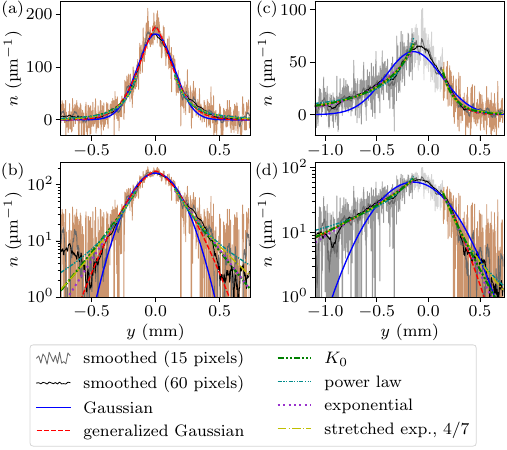}
    \caption{Analysis of density profiles for expansion in symmetric (a,~b) and asymmetric (c,~d) disorder. 
    Several expectations for density profiles have been fitted to the respective data, see legend on bottom. Additionally, smoothed data computed by convolutions with box functions of the given widths are shown for enhanced visibility. Note that both (a,~b) and (c,~d) display roughly 800 camera pixels along $y$. 
    (a)~Data from $\disorderstrength \approx \SI{642}{\nano K} \times \boltzmann$ and $t \approx \SI{0.82}{\second}$. For both the zeroth-order Hankel function $K_0$~\cite{shapiro_cold_2012, shapiro_expansion_2007} and the exponential function, the region $|y| < \SI{0.15}{\milli\meter}$ was excluded. The exponential function was fitted independently for both directions $\pm y$. 
    (b)~Same as (a) but with logarithmic density scale. 
    (c)~Data from $\disorderstrength \approx \SI{691}{\nano K} \times \boltzmann$ and $t = \SI{1}{\second}$. All shown functions were fitted independently for both directions $\pm y$. Segmentation along $y$ is identical to Figs.~\ref{fig5} and ~\ref{fig6}. 
    (d)~Same as (c) but with logarithmic density scale. 
    }
    \label{fig10}
\end{figure}

In this section, we compare several expectations of density profiles fitted to the recorded line densities $n(y)$. 
For that, we use two representative examples. 
The first, shown in Fig.~\ref{fig10}(a,~b), is from the expansion in strong symmetrical disorder, see Sec.~\ref{subsec:symmetric_disorder}. 
The second, shown in Fig.~\ref{fig10}(c,~d), is from the expansion in strong asymmetrical disorder, see Sec.~\ref{subsec:symmetric_disorder}. 
Both images were taken for an expansion time of $t\approx \SI{1}{\second}$ and have been averaged from 50 realizations, see Sec.~\ref{sec:experimental_methods} for more details. 
We plot each image in both linear and logarithmic density scales to emphasize the exponential character of the different density profiles or the lack thereof. 
Additionally, we show our data after smoothing it via convolution with a top-hat kernel, see gray lines. 
For the width of the smoothing kernel, we use both 15 and 60 pixels. 
Note that we only show the smoothed line plots but do not evaluate them. 
All shown fits are performed to the recorded densities. 
In the following, we discuss the choice of density profiles. 

To begin with, we fit a Gaussian function, see Eq.~\eqref{eq:gauss_wavefunction}, as it is the basis for all observables described in Sec.~\ref{sec:diffusion_observables} and is generally a good albeit naive approximation for our density profile. 
For a Bose-Einstein condensate (BEC) undergoing diffusion, Refs.~\cite{shapiro_expansion_2007, shapiro_cold_2012} derive a zeroth-order modified Hankel function $K_0$ and, as Ref.~\cite{shapiro_cold_2012} states, expansion of a degenerate Fermi gas is not expected to look very different. 
Therefore, we fit $K_0$ to the distribution tails. 
Note that this function is approximated well by an exponential function for a large distance to the origin, meaning a purely diffusive, delocalized BEC is already expected to exhibit exponential tails. 
Further, as it is the general expectation for an Anderson-localized density profile, we fit an exponential function. 
Contrarily, Ref.~\cite{muller_comment_2014} claims a stretched-exponential function as the density profile for the case of diffusive spread with an energy-dependent diffusion coefficient. 
Note that we use the same exponent of 4/7 for the fits presented here. 
The authors of Ref.~\cite{muller_comment_2014} further add that the tails should follow a power law in the localization scenario. 
Finally, we fit a generalized Gaussian, see Eq.~\eqref{eq:generalized gauss}. 
This function includes an additional exponent $\nu_\mathrm{gG}$ in the argument compared to the Gaussian and can be used to indicate the transition between a normal and exponential distribution~\cite{roati_anderson_2008,Chen2010}. 

As can be seen in Fig.~\ref{fig10}, all density distributions are, in principle, compatible with the experimental data due to the large noise. 
For the symmetric disorder, see Sec.~\ref{subsec:symmetric_disorder}, the generalized Gaussian appears (exponent of $\nu_\mathrm{gG} \approx 1.50$) closest to the data, which is expected as it has the largest amount of free parameters. 
The Gaussian function underestimates the tails slightly, while most other functions describe it relatively well. 
For completeness' sake, the power-law fit yields an exponent of $\nu_\mathrm{pl} \approx -2.63$ ($\nu_\mathrm{pl} \approx -2.85$) for the tail toward the $+y$ ($-y$) direction. 

For the asymmetric disorder, which is displaced toward the $+y$ direction, see Sec.~\ref{subsec:asymmetric_disorder}, the Gaussian function exhibits the largest discrepancy. 
However, all other functions describe the outer regions (dark gray and brown lines outside $y=0$) very well. 
Here, the generalized Gaussian yields an exponent of $\nu_\mathrm{gG} \approx 1.29$ ($\nu_\mathrm{gG} \approx 0.76$) for the $+y$ ($-y$) direction, while the power-law exponent is found to fit best at $\nu_\mathrm{pl} \approx -2.71$ ($\nu_\mathrm{pl} \approx -1.20$). 

We conclude that the low signal-to-noise ratio does not allow for a reliable analysis of the distribution shape in general, but especially the tails. 
In fact, this was the motivation for the careful investigation of subdiffusion.

\section{Localized fraction}
\label{app:floc}

\begin{figure}
    \centering
    \includegraphics[width=86mm]{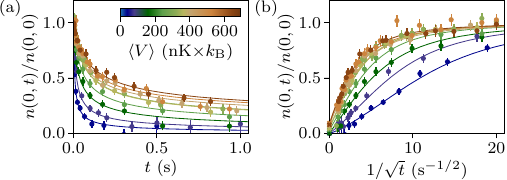}
    \caption{Determining the localized fraction $f_\mathrm{loc}$. 
    (a)~Relative peak density $n(t) / n(0)$ (circles) over expansion time $t$. Lines are fits with the anomalous-diffusion model Eq.~\eqref{eq:floc} where $f_\mathrm{loc}$, the value for $t \rightarrow \infty$ is the only free parameter. Errors are calculated from error propagation. 
    (b)~Plotting $n(t) / n(0)$ over the inverse square root of time visualizes $f_\mathrm{loc}$ as the density-axis intercept (crosses). 
    }
\label{fig11}
\end{figure}

As introduced in Ref.~\cite{jendrzejewski_three-dimensional_2012}, the localized fraction estimates the portion of atoms at the cloud peak that would not have diffused away due to being localized, 
\begin{equation}
\label{eq:floc_in_general}
    f_\mathrm{loc} = \lim_{t\to\infty} \frac{n(0, t)}{n(0, 0)}. 
\end{equation}
We modified its computation to fit our expansion along a single dimension and implemented the full anomalous-diffusion power law as in Eq.~\eqref{eq:anomalous_diffusion} by using the model
\begin{equation}
\label{eq:floc}
    \frac{n(0, t)}{n(0, 0)} = f_\mathrm{loc} + (1 - f_\mathrm{loc}) \sqrt{\frac{\sigma^2(0)}{2 D_\alpha t^\alpha + \sigma^2(0)}}, 
\end{equation}
where we fix the diffusion exponent $\alpha$ and coefficient $D_\alpha$ to the values we extract as described in Sec.~\ref{sec:diffusion_observables} and use $\sigma(0) = \SI{53}{\micro\meter}$. 
For the relative peak density $n(0, t) / n(0, 0)$, we use the approximation of $n(t) \approx w(t) - w_\mathrm{noise}$ mentioned in Sec.~\ref{subsec:ipw}, multiplied by the factor $N(0) / N(t)$ to compensate for atom losses. 
With that, $f_\mathrm{loc}$ is extracted as the only free parameter from fitting the right side of Eq.~\eqref{eq:floc}, see lines in Fig.~\ref{fig11}. \\

\end{document}